\documentclass[aps,epsfigure,twocolumn,superscriptaddress,showkeys,nofootinbib]{revtex4-1}
\usepackage[colorlinks=true,linkcolor=blue,urlcolor=blue,citecolor=blue,pdfusetitle]{hyperref}
\usepackage[utf8]{inputenc}
\usepackage[english]{babel}
\usepackage{amsmath}
\usepackage[caption = false]{subfig}
\usepackage{graphicx,epstopdf}
\usepackage{blindtext}
\usepackage[table,xcdraw]{xcolor}
\usepackage{lipsum}
\usepackage{amsfonts}
\usepackage{bbm}
\usepackage{amssymb}
\usepackage{enumerate}
\usepackage{comment}
\usepackage{color}
\usepackage{latexsym}
\usepackage{multirow}
\usepackage{threeparttable}
\usepackage{hyperref}

\usepackage{times,txfonts}
\usepackage[normalem]{ulem}
\graphicspath{ {./figures/} }
\usepackage{subcaption}
\usepackage{algorithm}
\usepackage{algpseudocode}
\usepackage{tikz}
\usetikzlibrary{quantikz}
\usepackage{booktabs}
\usepackage{ragged2e}
\usepackage{comment}
\usepackage{orcidlink}

\newcommand\lqg[1]{{\color{black} #1}}
\newcommand\lqgrev[1]{{\color{black} #1}}
\newcommand\djgs[1]{{\color{black} #1}}

\usepackage[most]{tcolorbox}
\tcbuselibrary{theorems,skins}

\newtcbtheorem[auto counter]{prop}{PEP}{%
  enhanced,
  colback=blue!4!white,
  colframe=blue!75!black,
  fonttitle=\bfseries,
  coltitle=white,
  colbacktitle=blue!75!black,
  sharp corners,
  boxrule=0.8pt,
  title style={left color=blue!100!black,right color=blue!80!black},
  separator sign=:\ %
}{PAP}

\begin{document}

\title{Neural network for excess noise estimation in continuous-variable quantum key distribution under composable finite-size security}

\author{Lucas Q. Galvão \orcidlink{0009-0001-2682-250X}}
\email{Corresponding author: Lucas Q. Galvão \newline Email: lqgalvao3@gmail.com}

\affiliation{QuIIN - Quantum Industrial Innovation, Centro de Competência Embrapii Cimatec. SENAI CIMATEC, Av. Orlando Gomes, 1845, Salvador, BA, Brazil CEP 41850-010}

\author{Davi Juvêncio G. de Sousa\orcidlink{0009-0009-1946-5573}}
\email{davi.juvencio@fbter.org.br}
\affiliation{QuIIN - Quantum Industrial Innovation, Centro de Competência Embrapii Cimatec. SENAI CIMATEC, Av. Orlando Gomes, 1845, Salvador, BA, Brazil CEP 41850-010}

\author{Micael Andrade Dias \orcidlink{0000-0001-6394-9174}}
\email{micael.dias@fieb.org.br}
\affiliation{Department of Electrical and Photonics Engineering, Technical University of Denmark, 2800  Lyngby, Denmark}
\affiliation{QuIIN - Quantum Industrial Innovation, Centro de Competência Embrapii Cimatec. SENAI CIMATEC, Av. Orlando Gomes, 1845, Salvador, BA, Brazil CEP 41850-010}

\author{Nelson Alves Ferreira Neto \orcidlink{0000-0003-2278-1082}}
\email{nelson.neto@fieb.org.br}
\affiliation{QuIIN - Quantum Industrial Innovation, Centro de Competência Embrapii Cimatec. SENAI CIMATEC, Av. Orlando Gomes, 1845, Salvador, BA, Brazil CEP 41850-010}

\begin{abstract}
Parameter estimation is a critical step in continuous-variable quantum key distribution (CV-QKD), as the statistical uncertainty from a finite data size leads to pessimistic worst-case bounds that drastically reduce the secret key rate and range. While machine learning techniques have been proposed for this task, they have lacked the rigorous statistical framework necessary for integration into a composable security proof. In this work, we bridge this gap by introducing a statistically rigorous framework for using neural networks for parameter estimation in CV-QKD with quantifiable composable security. We develop a neural network estimator for the excess noise and, crucially, derive its worst-case confidence interval using a delta method approach, ensuring the estimation fails with a probability not exceeding $\epsilon_{\rm PE}$. This allows the network to be integrated into a parameter estimation protocol that is operationally equivalent to the standard maximum likelihood method but yields significantly tighter parameter bounds. Our numerical results demonstrate that this method provides substantially more precise estimates, which directly translates into a higher secret key rate and extended transmission distance over a fiber channel under a collective Gaussian attack. This work establishes that machine learning can be securely and effectively harnessed to overcome a key performance limitation in practical CV-QKD systems.

\end{abstract}

\keywords{quantum cryptography; continuous-variable quantum key distribution; parameter estimation; neural networks.}
\maketitle

\section{Introduction}

Security in communication is a fundamental aspect of contemporary society, as it enables the sharing of sensitive information without the risk of potential leaks \cite{10.1145/359168.359176, un2024globalprinciples, enisa2024, Pirandola:20}. However, the emergence of recent algorithms has threatened this security, such as the quantum algorithm for factoring integers in logarithmic time \cite{Shor1994, Shor1999}. Considering these challenges, efforts in the field of quantum communication have been made to use properties inherent to quantum physics to ensure unconditionally secure communication \cite{RevModPhys.74.145, renner2006securityquantumkeydistribution, RevModPhys.94.025008, RevModPhys.81.1301}. In this context, the use of continuous-variable quantum key distribution (CV-QKD) emerges as a potential alternative, since it has greater adaptability to the current components found in coherent optical telecommunications systems \cite{usenko2025continuousvariablequantumcommunication, 10.1063/5.0179566, Laudenbach2018, e17096072}. 

In a generic CV-QKD protocol, quantum information can be encoded onto coherent states by modulating the amplitude and phase quadratures of laser light, typically using electro-optical modulators at the transmitter to establish a secret key between two legitimate QKD users (Alice and Bob) \cite{PhysRevLett.88.057902, Grosshans2003b, Weedbrook2004}. These states are transmitted through a quantum channel that is assumed to be fully under the control of a potential eavesdropper, conventionally referred to as Eve. The security of CV-QKD protocols employing Gaussian-modulated coherent states was initially proven in the asymptotic limit~\cite{PhysRevA.71.052304, PhysRevLett.97.190502, Garcia2006} and later extended to the finite-size regime, ensuring universal composability against both collective~\cite{Leverrier2010} and general coherent attacks~\cite{PhysRevLett.109.100502}. Beyond continuous modulation, discrete modulation of coherent states has also been extensively studied and shown to offer promising performance and security guarantees~\cite{PhysRevLett.102.180504, Denys2021explicitasymptotic, Bauml2024securityofdiscrete}.

Since the channel is under the control of Eve, the smooth min-entropy has to be bounded by the worst case compatible with the observed measurement data \cite{Devetak2005}. Thus, a fundamental procedure in CV-QKD is the estimation of the channel parameters, such as the transmittance $T$ and the excess noise $\xi$ \cite{usenko2025continuousvariablequantumcommunication, 10.1063/5.0179566, Laudenbach2018, e17096072}. In principle, there are other parameters to be estimated, but transmittance and excess noise have the most significant impact on the secret-key rate, where the latter has a drastic impact for long distances \cite{Huang2016, Laudenbach2018, doi:10.1126/sciadv.adi9474}. For a finite-key security analysis, parameter estimation must ensure that the key is secure against any eavesdropper attack, up to a probability of failure \cite{Scarani2008, Cai_2009}. 

There is broad consensus in the literature that the maximum likelihood estimation (MLE) method offers strong security guarantees for the protocol \cite{PhysRevA.93.042343, PhysRevResearch.6.023321, PhysRevA.96.042332, PhysRevLett.125.010502}, as its confidence intervals can be explicitly computed and depend directly on the chosen significance level $\epsilon_{PE}$ \cite{monfort1982cours}. In fact, the first finite-key analysis under the assumption of collective Gaussian attacks was provided in ref. \cite{Leverrier2010} using MLE. Recently, machine learning techniques have been introduced for various tasks in CV-QKD \cite{info14100553}, including parameter estimation \cite{PhysRevA.97.022316, Chin2021, Luo_2022}. However, most of these works do not account for the probability of estimation failure when neural networks are employed. Consequently, there is currently no established statistical security framework for using neural networks within the CV-QKD context.

In this work, we provide a finite-size security analysis demonstrating that neural networks can be reliably used for CV-QKD with quantifiable failure probabilities $\epsilon_{PE}$, which possesses an operational interpretation and composability security. We demonstrated our analysis in a neural network architecture using a parameter estimation protocol (PEP) operationally equivalent to the standard method presented in ref. \cite{Leverrier2010}. Our results showed that neural networks can provide more precise estimations in order to gain more distances without compromising the security of the protocol. 

\section{Protocol and model definition} \label{sec:protocol}

In this work, we will consider the coherent-state protocol with Gaussian modulation \cite{GG02}. The protocol starts with Alice preparing $N$ displaced vacuum states $\lvert q_i + i p_i \rangle$ by modulating both the amplitude and phase quadratures. The displacements $q_i$ and $p_i$ are independent random variables drawn from the normal distribution $\mathcal{N}(0,V_A)$. These states are transmitted over an untrusted channel with transmission $T$ and excess noise $\xi$, assumed to be under Eve’s control. Upon reception, Bob performs homodyne detection by switching randomly between the phase and amplitude quadratures. Here, Alice’s use of modulation to prepare these states constitutes a prepare-and-measure protocol, known to have an entanglement-based equivalent \cite{Grosshans2003}, which
is used for the security analysis \cite{Garcia2006}.

For Gaussian channels, the relationship between the signal sent by Alice and the signal received by Bob is given by the linear model
\begin{equation} \label{eq:channel}
    y_i = tx_i + z_i \,,
\end{equation}
where $\{y_i\}_N$ and $\{x_i\}_N$ are the classical data related to the random variables of Bob and Alice, respectively. In the measurement, Alice's signal is affected by the parameter $t=\sqrt{T}$ and the noise $\{z_i\}_N$, represented by a random variable with zero mean and variance $\sigma^2 = \mu + t^2 \xi$. The parameter $\mu$ is the quantum duty (“qu-duty”) associated with detection:  $\mu=1$ for homodyne and $\mu = 2$ for heterodyne \cite{Pirandola2021}. Operationally, the transmission can be evaluated with
\begin{equation}
    T \approx \eta_{\text{eff}} 10^{-0.02 \ell}
\end{equation}
where $\eta_{\text{eff}}$ is the known quantum efficiency of Bob’s detection and $\ell$ is the distance in kilometers between Alice and Bob. In this case, we assume an optical fiber with losses of $0.2$ dB per kilometer. More generally, we consider the possibility that Eve can access side-channel information resulting from imperfections in the detection setup. In this analysis, we focus on the receiver’s detection inefficiency and assume that the fraction $1 - \eta_{\text{eff}}$ of the incoming photons that are not detected is not simply lost to the environment, but is instead collected by Eve and incorporated into her attack strategy \cite{Pirandola2021}.

In the context of QKD, the objective is to extract a positive secret key rate while guaranteeing composable security considering the signals $\{y_i\}_N$ and $\{x_i\}_N$ in the presence of channel loss and excess noise \cite{Scarani2009, RENNER2008, lo2014secure}. 
A widely adopted method for evaluating this quantity is the Devetak–Winter bound~\cite{Devetak2005}

\begin{equation} \label{eq:dw}
    I(x:y) - \sup_{\mathcal{N}: A' \rightarrow B} \chi(y:E)
\end{equation}
where $I(x:y)$ denotes the mutual information between Alice’s and Bob’s classical variables $x$ and $y$  \cite{cover1999elements}, while $\chi(y:E)$ represents the Holevo information between Bob’s variable $y$ and the adversary’s quantum system $E$ \cite{nielsen2010quantum}. The supremum is taken over all channels $\mathcal{N}: A \rightarrow B$ that are consistent with the statistics observed by Alice and Bob during the parameter estimation. For the entanglement-based protocol \cite{Grosshans2003}, it can be described by the covariance matrix

\begin{equation} \label{eq:mat_cov}
    \Gamma =  \begin{pmatrix}
(V_A + 1)\mathbb{I}_2 &  t Z \sigma_z\ \\
t Z \sigma_z &  (t^2V_A + \sigma^2)\mathbb{I}_2 \\
\end{pmatrix} \,,
\end{equation}
where $\sigma_z$ is the Pauli matrix and $Z = \sqrt{V_A^2 + 2V_A}$ for Gaussian modulation \cite{Laudenbach2018}.

Using the covariance matrix, $\chi(y: E)$ can be determined by its symplectic eigenvalues \cite{Laudenbach2018}. Following the quantum stage, classical data processing and a mathematically rigorous security analysis are performed to distill a secret key of certified length \cite{Pirandola2021}. However, Eq \eqref{eq:dw} does not take into account the effects of post-processing data after Bob's measurements. For example, the parameter estimation significantly impacts this value because $m$ signals are used for this estimation, reducing the total number of data used to generate the raw key to $n \equiv N - m$ \cite{Cai_2009}. Also, the reconciliation efficiency $\beta$ is another important value to be considered, since it estimates the amount of information Bob can recover from Alice, limiting the mutual information \cite{Almeida2023}.

Furthermore, since the finite number of quantum states exchanged by Alice and Bob inevitably reduces the achievable key length, incorporating finite-size effects is indispensable to guarantee composable security \cite{cryptoeprint:2000/067} up to a failure probability $\epsilon_{PE}$, thereby ensuring the protocol remains operationally meaningful in realistic conditions \cite{Jain2022}. This can be quantified considering the statistical error present in post-processing steps \cite{Leverrier2010, Pirandola2021}, characterized by
\begin{equation}
    \epsilon = p_{\text{ec}} \epsilon_{\mathrm{PE}} + \epsilon_{\text{cor}} + \epsilon_{\text{sec}} \,
\end{equation}
where $\epsilon_{\text{cor}}$ and $\epsilon_{\text{sec}}$ indicate that the protocol satisfies $\epsilon$-correctness and $\epsilon$-secrecy, respectively. The secrecy parameter $\epsilon_{\text{sec}}$ can be further decomposed as $\epsilon_{\text{sec}} = \bar{\epsilon} + \epsilon_{\mathrm{PA}}$, where $\bar{\epsilon}$ is the smoothing parameter and $\epsilon_{\mathrm{PA}}$ denotes the probability of failure of the privacy amplification step. Additionally, $p_{\text{ec}}\epsilon_{PE}$ accounts for the probability of failure in parameter estimation, while $p_{\text{ec}} = 1 - \text{FER}$ represents the probability of successful error correction, with $\text{FER}$ being the frame error rate \cite{PhysRevLett.109.100502, Pirandola2021}.

The parameter $\epsilon$ must be composable and have an operational interpretation in order to ensure that it meets the security requirements \cite{renner2006securityquantumkeydistribution}. In the context of DV-QKD, ref. ~\cite{Scarani2008} showed that $\epsilon$ satisfies the composability criterion and corresponds to the maximum failure probability of the protocol, meaning the maximum probability that an eavesdropper obtains non-negligible information about the final key $k_{\epsilon}$. This notion was later extended to CV-QKD in ref. ~\cite{Leverrier2010}. Specifically, the parameter $\epsilon_{\mathrm{PE}}$ affects the estimation of the covariance matrix in Eq.~\eqref{eq:mat_cov}, requiring the replacement of the Holevo information $\chi(y: E)$ by its smooth version $\chi_{\epsilon_{\mathrm{PE}}}(y: E)$. The parameters $\epsilon_{\mathrm{PA}}$ and $\bar{\epsilon}$ enter the secret key rate expression via the finite-size correction term $\Delta(n)$, which adjusts the asymptotic key rate to account for statistical fluctuations and composable security requirements \cite{renner2006securityquantumkeydistribution}. Thus, the secret key rate is finally written as

\begin{equation} \label{eq:skr_finite}
    k_{\epsilon} = \frac{np_{EC}}{N}(\beta I(x:y) - \chi_{\epsilon_{PE}}(y:E) - \Delta (n))
\end{equation}
and the last term is explicitly defined as

\begin{equation}
    \Delta (n) \equiv 4 \log_2 (\sqrt d + 2)\sqrt{\frac{1}{n}\log_2 \left (\frac{18}{p_{\mathrm{ec}}^2\epsilon_s^4} \right )}  + \frac{2}{n}\log_2(1/\epsilon_{PA}) \,,
\end{equation}
where $d$ denotes the number of bits per quadrature used during discretization (e.g., $d = 2^5=32$ for a 5-bit discretization), $\bar\epsilon$ is the smoothing parameter, and $\epsilon_{\mathrm{PA}}$ is the failure probability of the privacy‐amplification procedure \cite{Leverrier2010}. Both $\bar\epsilon$ and $\epsilon_{\mathrm{PA}}$ are intermediate quantities to be optimized numerically. The first term of $\Delta(n)$, namely the square‐root term, quantifies the convergence rate of the smooth min‐entropy — the relevant metric for key length — of an i.i.d.\ state under collective attacks toward its von Neumann entropy, since only in the asymptotic limit does the smooth min‐entropy equal the von Neumann entropy. Its derivation was done in ref. \cite{Pirandola2021}, which used the framework proposed in ref. \cite{tomamichel2013frameworknonasymptoticquantuminformation}. The second term directly reflects the security contribution of the failure probability $\epsilon_{\mathrm{PA}}$ in the privacy‐amplification step. 

\section{Parameter estimation effects on protocol security}

The parameter estimation is, without any doubt, the main problem for CV-QKD in finite-size scenario: The uncertainty related to the estimation limits the secret key-rate, since one can never estimate a secret key-rate below its real value \cite{Leverrier2010}. From the covariance matrix in Eq. \eqref{eq:mat_cov}, one can see that we need to compute both the transmittance and excess noise values. In this post-processing pipeline, we consider that Alice computes the covariance matrix with the data Bob publishes on the authenticated channel, so the variables $V_A$ and $Z$ are not considered problematic in the parameter estimation stage.

The main problem is that Bob and Alice do not know these parameters, since it is assumed that the channel can be freely controlled by Eve. The law of large numbers guarantees that when $m \rightarrow \infty$ $\mathbb{E}[\hat t] \equiv t$ and $\mathbb{E}[\hat \sigma^2] \equiv \sigma^2$  \cite{casella2024statistical}, so there is no need for error analysis in asymptotic limit. In practical implementations, it is obviously impossible to achieve this result, such that one needs to consider the  probability of failure of the parameter estimation $\epsilon_{PE}$ using statistical analysis for the estimators. 

The maximum likelihood estimation is widely recognized as the standard method in the field, since it is compatible with statistical analyses considering the confidence interval \cite{PhysRevA.93.042343, PhysRevResearch.6.023321, PhysRevA.96.042332, PhysRevLett.125.010502}. For the linear model, 
\begin{equation} \label{eq:MLE}
    \hat t = \sum_i^m \frac{y_i x_i}{x_i ^2} \, \quad \text{and} \, \quad \hat \sigma^2 = \sum_i^m \frac{(y_i - \hat tx_i)^2}{m} \,.
\end{equation}

The confidence interval is then computed considering the lower bound for $t_{min}$ and the upper bound for $\sigma^2_{max}$:

\begin{equation} \label{eq:t_min}
    t_{min-\text{MLE}} \approx \hat t_{\text{MLE}} - z_{\epsilon_{PE}/2} \sqrt{\frac{\hat \sigma^2}{m V_A}} \,,
\end{equation}

\begin{equation} \label{eq:sigma_max}
  \sigma^2_{\max-\text{MLE}} \approx \hat \sigma^2_{MLE} + z_{\epsilon_{PE}/2} \frac{\hat \sigma^2 \sqrt{2}}{\sqrt{m}} \,,
\end{equation}
where $z_{\epsilon_{PE}/2} = \text{erf}^{-1}(1 - \epsilon_{PE}/2)$ and $\text{erf}(x)$ is the error function, \lqg{ensuring that the key is not overestimated by one of the authenticated parties with probability at least $1 - \epsilon_{PE}/2$, i.e., 

\begin{equation}
    P(\sigma^2 \leq \sigma_{\mathrm{max}}^2) \geq 1 - \epsilon_{\mathrm{PE}}/2 \quad
\end{equation}

and

\begin{equation}
    P(t \geq t_{\mathrm{min}}) \geq 1 - \epsilon_{\mathrm{PE}}/2 \,.
\end{equation}}

Definition \eqref{eq:t_min} and \eqref{eq:sigma_max} guarantees that we never estimate a transmittance higher than its real value or a noise variance lower than its real value, except with probability $\epsilon_{PE}/2$. Thus, it holds both composability and operational interpretation.

In this context, the covariance matrix assuming the probability of failure of MLE is given by
\begin{equation} \label{eq:cov_mat_worst}
    \Gamma_{\epsilon_{PE}} =  \begin{pmatrix}
(V_A + 1)\mathbb{I}_2 &  t_{min} Z \sigma_z\ \\
t_{min} Z \sigma_z &  (t_{min}^2V_A + \sigma^2_{max})\mathbb{I}_2 \\
\end{pmatrix} \,,
\end{equation}
which means that there exists a confidence set $\mathcal{C}_{\epsilon_{PE}}$ such that the covariance matrix $\Gamma_{\epsilon_{PE}}$ lies within $\mathcal{C}_{\epsilon_{PE}}$ with probability at least $1 - \epsilon_{PE}/2$. Thus, the secret key rate that accounts for the probability of failure in parameter estimation can be computed using PEP~\ref{PAP:2}.

\begin{prop}{Parameter estimation via maximum likelihood estimation in the finite-size scenario} {2}
    \begin{enumerate}
    \item Since Alice only has access to {her transmitted} signals $y$, Bob needs to broadcast $m$ signals over an authenticated channel so that {Alice} can estimate $t$ and $\sigma^2$.
    \item Alice uses estimator from Eq. \eqref{eq:MLE} to estimate $t$ and $\sigma^2$, using the $m$ correlated data.
    \item Alice uses the statistical analysis from Eq. \eqref{eq:t_min} to compute $t_{min}$ and from Eq. \eqref{eq:sigma_max} to compute $\sigma^2_{max}$. \label{step:PEP_s3}
    \item Alice uses these results to write the covariance matrix from Eq. \eqref{eq:cov_mat_worst} and, finally, compute $\chi_{\epsilon_{PE}}(y: E)$.
\end{enumerate}
\end{prop}

Note that step \ref{step:PEP_s3} becomes redundant in the asymptotic limit. 

\section{Worst-case confidence intervals for neural networks}
\label{Worst-case}

The computational modeling of systems with output $Y$ is described by a function $f(\mathrm{X}, \theta^*)$, where $\theta^*$ denotes the parameters of the model. The output is assumed to be affected by an additive error term \(\varepsilon\), which is independently and identically distributed according to a normal distribution \(\varepsilon \sim \mathcal{N}(0, \sigma^2_\varepsilon)\). For each observation \(i = 1, 2, \ldots, N\), the model is represented as
\begin{equation} \label{eq:system}
    \mathrm{Y}_i = f(\mathrm{X}_i, \theta^*) + \varepsilon_i \,,
\end{equation}
where $\mathrm{X}_i$ denotes the input corresponding to the \(i\)-th observation \cite{seber2005nonlinear}. This modeling framework is well established in computational learning theory, as it enables statistical generalizations across a broad class of inference tasks \cite{Barron1994, Papadopoulos2001}. \lqg{While statistical estimators derives confidence intervals directly from the known sampling distribution of the physical parameter estimators \cite{casella2024statistical}, machine learning-based estimation includes the explicit error term $\varepsilon$, as it accounts for the approximation error between the learning-theoretic framework's predictions and the true underlying function.}

The central challenge, therefore, lies in demonstrating that neural networks can be effectively described within this framework under appropriate assumptions, thereby ensuring reliable and secure parameter estimation. In this work, we adopt such a perspective, drawing inspiration from the delta method outlined in refs. \cite{Chryssolouris1996, Hwang1997}.

In general, neural networks for prediction give an output 
\begin{equation}
    \hat{\mathrm{Y}} = f(\mathrm{X}_i, \hat \theta) \,,
\end{equation}
which can approximate from Eq. \eqref{eq:system} by minimizing the error function
\begin{equation} \label{eq:min_loss}
    S(\theta) = \sum_{i=-1}^N [\hat{\mathrm{Y}}_i - f(\mathrm{X}_i; \theta^*)]^2 \,.
\end{equation}

This procedure is expected to bring $\hat \theta$ closer to $\theta^*$, \lqg{where $\theta^{*}$ denotes the optimal set of network parameters that minimizes the expected prediction error and $\hat{\theta}$ is the set obtained by minimizing the empirical loss}.  In principle, minimizing the loss function corresponds to finding the optimal set of network parameters that best approximates the underlying functional relationship between the inputs and the outputs \cite{NEURIPS2018_a41b3bb3}. Given a sufficiently expressive architecture and representative training data, the neural network learns a map that minimizes the discrepancy between the predicted and true values of the target variable.

Thus, a first-order Taylor expansion can be employed to approximate $f(\mathrm{X}_i, \theta^*)$ from $f(\mathrm{X}_i, \hat \theta)$, represented as
\begin{equation} \label{eq:first_Taylor}
    f (\mathrm{X}_i; \hat \theta) \approx f(\mathrm{X}_i, \theta^*) + \mathbf{f}_0^T \cdot (\hat \theta - \theta^*)\,,
\end{equation}
where 
\begin{equation}
    \mathbf{f}_0^T = \left ( \frac{\partial f(\mathrm{X}_i, \theta^*)}{\partial\theta_1^*}, \frac{\partial f(\mathrm{X}_i, \theta^*)}{\partial\theta_2^*}, \ldots, \frac{\partial f(\mathrm{X}_i, \theta^*)}{\partial\theta_p^*}\right ) \,
\end{equation}
with the subscript ``$0$" indicating the set of points that are not used in the least-squares estimation of $\theta ^*$. In this sense, the difference between the real and predicted value is written as
\begin{equation}
    \mathrm{Y}_0 - \hat{\mathrm{Y}}_0 \approx \varepsilon_0 -  \mathbf{f}_0^T  \cdot (\hat \theta - \theta^*) 
\end{equation}

The first term corresponds to the intrinsic measurement noise, while the second term captures the uncertainty in the model prediction due to the estimation error in the parameters. \lqg{The first-order Taylor expansion in Eq.~\eqref{eq:first_Taylor} introduces a linearization error of order \( \|\hat{\theta} - \theta^{*}\|^2 \). This error becomes negligible in the large-sample regime where \(\hat{\theta}\) converges to \(\theta^{*}\). Moreover, the confidence intervals derived via the delta method are conservative by construction, meaning they account for such higher-order uncertainties and preserve the composable security of the estimation scheme.}

\lqg{In this framework, we introduce the hypothesis that the prediction error $\varepsilon_0$ follows a zero-mean Gaussian distribution based on two complementary considerations: the physical channel model and the statistical properties of the trained neural network. First, because the network is trained on data generated from the normal linear model, the resulting prediction errors are expected to approximately inherit Gaussian characteristics. Second, minimization of the MSE loss yields parameter estimates $\hat{\theta}$ that converge to a Gaussian random vector around its optimal value $\theta^*$. This follows from the fact that the loss gradient is a sum of many independent terms, inducing asymptotic normality via the central limit theorem.} Considering this approach, the parameter estimation error $\hat{\theta} - \theta^*$ can be approximated as following a multivariate normal distribution 
\begin{equation}
    \hat{\theta} - \theta^* \sim \mathcal{N}_p\left(0, \sigma^2_\varepsilon \left[\mathbf{F}^T(\hat{\theta}) \mathbf{F}(\hat{\theta})\right]^{-1} \right),    
\end{equation}
where $\mathbf{F}(\hat{\theta})$ denotes the Jacobian matrix of first-order partial derivatives of the model function $f(\mathrm{X}, \theta)$ with respect to the parameters \cite{Chryssolouris1996}, evaluated at $\hat{\theta}$:

\begin{equation} \label{eq:jac}
\mathbf{F}(\hat{\theta}) = \frac{\partial f(\boldsymbol{\mathrm{X}}, \hat{\theta})}{\partial \hat{\theta}} =
\begin{bmatrix}
\frac{\partial f_1(\mathrm{X}_1, \hat{\theta})}{\partial \hat{\theta}_1} & \frac{\partial f_1(\mathrm{X}_1, \hat{\theta})}{\partial \hat{\theta}_2} & \cdots & \frac{\partial f_1(\mathrm{X}_1, \hat{\theta})}{\partial \hat{\theta}_p} \\
\frac{\partial f_2(\mathrm{X}_2, \hat{\theta})}{\partial \hat{\theta}_1} & \frac{\partial f_2(\mathrm{X}_2, \hat{\theta})}{\partial \hat{\theta}_2} & \cdots & \frac{\partial f_2(\mathrm{X}_2, \hat{\theta})}{\partial \hat{\theta}_p} \\
\vdots & \vdots & \ddots & \vdots \\
\frac{\partial f_n(\mathrm{X}_n, \hat{\theta})}{\partial \hat{\theta}_1} & \frac{\partial f_n(\mathrm{X}_n, \hat{\theta})}{\partial \hat{\theta}_2} & \cdots & \frac{\partial f_n(\mathrm{X}_n, \hat{\theta})}{\partial \hat{\theta}_p}
\end{bmatrix}
\end{equation}

This formulation reflects how parameter uncertainty contributes to the overall variance of the model’s prediction, resulting in

\begin{equation}
    \mathrm{var}[\mathrm{Y}_0 - \hat{\mathrm{Y}}_0] \approx \sigma^2_\varepsilon + \sigma^2_\varepsilon \mathbf{f}_0^T (\mathbf{F}^T \mathbf{F})^{-1} \mathbf{f}_0 \,.
\end{equation}

The matrix $\mathbf{F}(\hat{\boldsymbol{\theta}})$ has dimensions $n \times p$, where $n$ is the number of samples used to estimate the parameters and $p$ is the number of parameters $\theta_i$ that compose the vector $\hat{\boldsymbol{\theta}}$. 

Once the neural network is trained, the variance of the additive error term $ \sigma^2_\varepsilon$ can be computed during the test process using an unbiased estimator based on the residual sum of squares:
\begin{equation} \label{eq:varS}
s^2 = \frac{ \left\| \mathbf{\mathrm{Y}} - \mathbf{f}(\boldsymbol{\mathrm{X}}, \hat{\theta}) \right\|^2 }{n - p} \,,
\end{equation}
which is constant for a given $\mathrm{Y} \in \mathcal{S}$, where $\mathcal{S}$ delimits the set of possible values for the excess noise estimation.  \lqgrev{Its value is computed from the same synthetic dataset used to evaluate the full Jacobian matrix \( \mathbf{F} \), while the Jacobian vector \( \mathbf{f}_0 \) corresponds to the gradient at the the actual input sample used to estimate the channel parameters \( \mathbf{X}_0 \). This separation allows the Jacobian matrix to be pre‑computed, while the lightweight vector \( \mathbf{f}_0 \) is evaluated per frame, ensuring statistical rigor through error propagation while maintaining real‑time feasibility.}

Then the variance estimator is used in the Student's $t$-distribution to quantify the uncertainty associated with the prediction $\hat{\mathrm{Y}}_0$. Thus, one can construct a confidence interval for the predicted value $\hat{\mathrm{Y}}_0$, which incorporates both the estimated variance and the sensitivity of the prediction to the parameters through the Jacobian vector $\mathbf{f}_0$:

\begin{equation} 
t_{n-p} \sim \frac{\mathrm{Y}_0 - \hat{\mathrm{Y}}_0}{\sqrt{\mathrm{var}[\mathrm{Y}_0 - \hat{\mathrm{Y}}_0]}} 
\approx \frac{\mathrm{Y}_0 - \hat{\mathrm{Y}}_0}{\sqrt{s^2 + s^2 \mathbf{f}_0^T (\mathbf{F}^T \mathbf{F})^{-1} \mathbf{f}_0}}
\end{equation}

\begin{equation*}
\approx \frac{\mathrm{Y}_0 - \hat{\mathrm{Y}}_0}{s \left(1 + \mathbf{f}_0^T (\mathbf{F}^T \mathbf{F})^{-1} \mathbf{f}_0 \right)^{1/2}}
\end{equation*}

and finally

\begin{equation} \label{eq:conf_int_nn}
\mathrm{Y}_0 - \hat{\mathrm{Y}}_0 \pm t_{n-p}^{\alpha/2} s \left(1 + \mathbf{f}_0^T (\mathbf{F}^T \mathbf{F})^{-1} \mathbf{f}_0 \right)^{1/2} \,.
\end{equation}

From a computational picture, the trained neural network is defined as a family of probability distributions on a sample space of excess noise $\mathcal{S}$, indexed by a parameter vector $\hat \theta \in \Theta$. Thus, the trained neural network acts as a statistical estimator $\hat{\theta}: \mathcal{S} \to \Theta$, approximating the mapping from data to channel parameters. The quantity in Eq.~\eqref{eq:sigma_max_nn} defines a confidence interval with probability $1 - \epsilon/2$:

\begin{equation} \label{eq:sigma_max_nn}
    \sigma^{2}_{max-NN} \approx \hat \sigma^2_{NN} +t_{n-p}^{\epsilon_{PE}/2}s(1 + \mathbf{f}_0^T\left[\mathbf{F}^T(\hat{\theta}) \mathbf{F}(\hat{\theta})\right]^{-1}\mathbf{f}_0) \,.
\end{equation}

Since neural networks can be computationally expensive, the efforts invested in them must be used on processes that have significant impacts on the key rate. As discussed, this is the case for excess noise \cite{Huang2016, Laudenbach2018, doi:10.1126/sciadv.adi9474}. In this case, the neural network to estimate the variance in the worst-case is then defined, and finally, we can estimate the excess noise via PEP \ref{PAP:3}. 

\begin{prop}{Parameter estimation via neural network in the finite-size scenario}{3}
    \begin{enumerate}
     \item Since Alice only has access to {her transmitted} signals $y$, Bob needs to broadcast $m$ signals over an authenticated channel so that {Alice} can estimate $t$ and $\sigma^2$.
    \item Alice uses estimator from Eq. \eqref{eq:MLE}  to compute $\hat t$ and a trained neural network to compute $\hat \sigma^2$.
    \item Alice uses the statistical analysis from Eq. \eqref{eq:t_min} to compute $t_{min}$ and from Eq. \eqref{eq:sigma_max_nn} to compute $\sigma^2_{max}$.
    \item Alice uses these results to write the covariance matrix from Eq. \eqref{eq:cov_mat_worst} and, finally, compute $\chi_{\epsilon_{\mathrm{PE}}}(y: E)$.
\end{enumerate}
\end{prop}

\begin{figure*}
    \centering
    \includegraphics[width=0.8\linewidth]{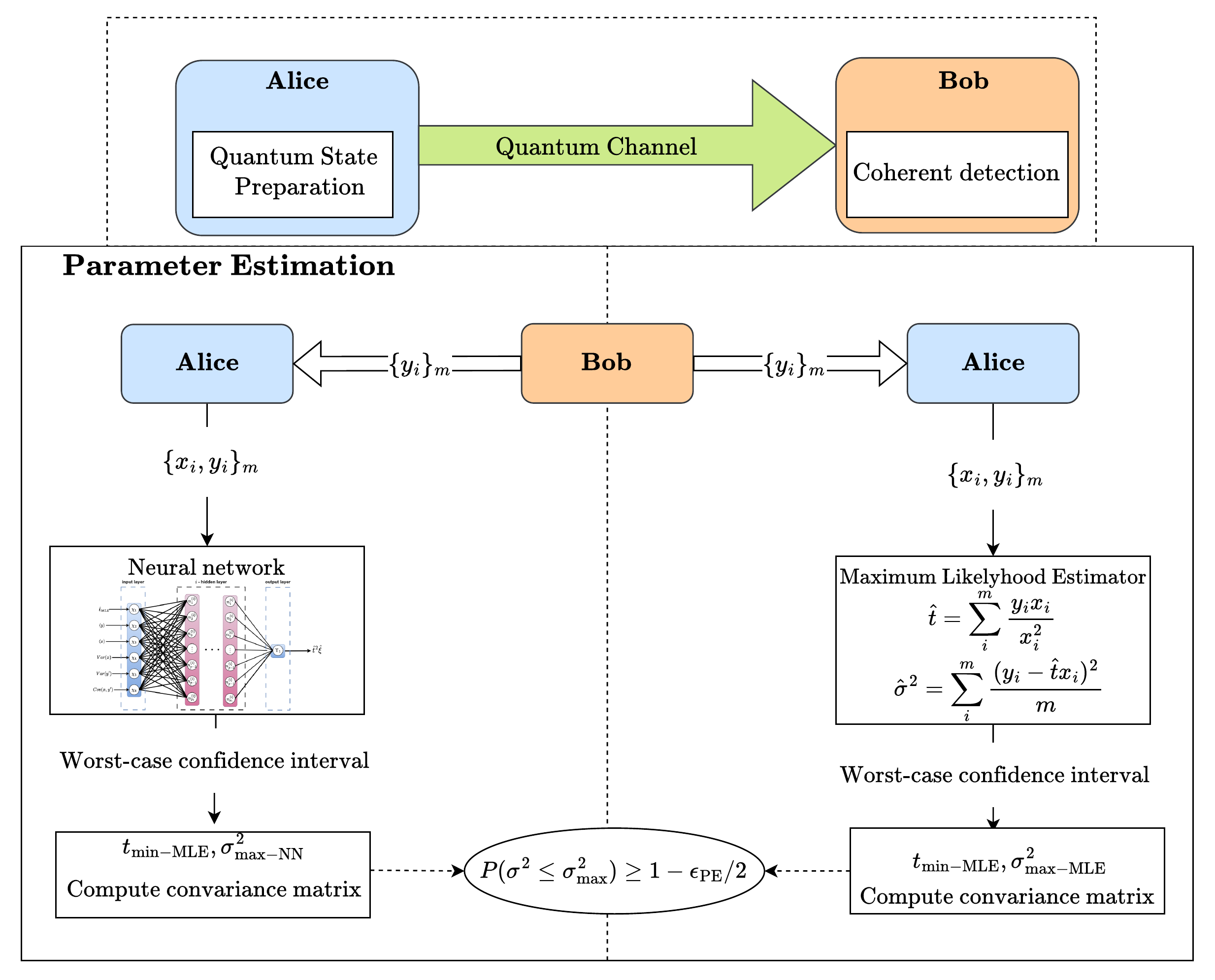}
    \caption{\justifying \lqg{Schematic representation of the parameter estimation step in a CV-QKD protocol. The two parallel branches illustrate the operational equivalence between the conventional MLE and the proposed neural-network estimator. In both branches, Alice and Bob use the same set of correlated quadrature measurements \(\{x_i, y_i\}_m\) to compute a worst-case upper bound \(\sigma^2_{\max}\) on the noise variance and a lower bound \(t_{\min\text{-MLE}}\) on the channel transmittance. These bounds are then used to construct the same worst-case covariance matrix \(\Gamma_{\epsilon_{PE}}\), from which the smooth Holevo information \(\chi_{\epsilon_{PE}}(y:E)\) entering the secret key rate is evaluated. }}
    \label{fig:scheme_PE}
\end{figure*}

\lqg{
The quantity in Eq. \eqref{eq:sigma_max_nn} defines a confidence interval with probability $1 - \epsilon_{PE}/2$ used in PEP \ref{PAP:3}, which is operationally equivalent to MLE-based parameter estimation method from PEP \ref{PAP:2}. This equivalence operates on two fundamental levels:

\begin{itemize}
    \item \textbf{Security Guarantee:} Both methods produce an upper bound $\sigma^2_{\text{max}}$ that satisfies the same probabilistic condition $P(\sigma^2 \leq \sigma^2_{\text{max}}) \geq 1 - \epsilon_{PE}/2$, which is the direct requirement for composable security in the parameter estimation stage.
    
    \item \textbf{Protocol Integration:} The value $\sigma^2_{\text{max}}$ generated by either method is inserted into the same worst-case covariance matrix (Eq.\ref{eq:cov_mat_worst}) and used to compute the same Holevo information quantity $\chi_{\epsilon_{PE}}(y:E)$ in the secret key rate (Eq.~\ref{eq:skr_finite}). From the perspective of the security protocol, the operation is indistinguishable, as shown in Fig. \ref{fig:scheme_PE}.
\end{itemize}
}

Therefore, once the network is trained based on the channel model, Alice can freely use it to perform estimation with just the $\{x\}_m$ and $\{y\}_m$ data, ensuring that $\Gamma_{\epsilon_{PE}} \in \mathcal{C_{\epsilon_{PE}}}$. An additional practical benefit is the possibility of training neural networks using synthetic data generated from known channel models. Since most realistic QKD channels can be well-approximated as Gaussian, this approach enables the use of pre-trained models during operation, eliminating the need for real-time training. This strategy reduces computational overhead while preserving the advantages in estimation precision, making neural networks a viable component in the implementation of efficient and secure QKD systems.

\section{Neural network model}
\label{NeuralNetworkModel}

In this section, we present the neural network architecture developed to implement the framework introduced previously. We detail the data input structure, the network's architectural design, and the training strategy employed. A discussion regarding the computational cost associated with the proposed neural network is provided in Appendix~\ref{ap:compcost}, \lqg{while the practical implementation for confidence interval is provided in Appendix ~\ref{ap:practical}.}

\subsection{Neural network inputs}

To estimate the noise variance parameter in a CV-QKD system, we designed a fully-connected feedforward neural network tailored to extract nonlinear correlations from statistical quadrature measurements of the channel. All input data are expressed in shot-noise units (SNU), and the network is trained to estimate the product $\hat{t}^2 \hat{\xi}$, since the parameter $\mu$ is constant for different frames. 

The input vector to the network is given by $\mathrm{X}_i \in \mathbb{R}^6$, composed of sufficient statistics computed from a sample of correlated variables $\{x_i, y_i\}_m$:
\begin{equation}
\label{eq:nnpe-inputs}
\mathrm{X}_i = \left\{ \hat{t}_{\text{MLE}},\ \langle x \rangle,\ \langle y \rangle,\ \text{Var}(x),\ a^2 (\text{Var}(y') - 1),\ \text{Cov}(x, y') \right\},
\end{equation}
where $\hat{t}_{\text{MLE}}$ is given by Eq. \eqref{eq:MLE}, and $y'$ is a preprocessed version of Bob’s variable $y$, defined as:
\begin{equation}
y'_i = y_i - \hat{t}_{\text{MLE}}x_i + \frac{\hat{t}_{\text{MLE}}}{a}x_i,
\end{equation}
with $a > 1$ representing an artificial amplification factor. This preprocessing step is designed to enhance the contribution of excess noise in the signal, making it more detectable by the neural network.

Under this transformation, the variance of $y'$ becomes:
\begin{equation}
\text{Var}(y') = (\hat{t} - \hat{t}_{\text{MLE}})^2 V_A + \frac{\hat{t}_{\text{MLE}}^2}{a^2}V_A + \hat{t}^2 \hat \xi + 1,
\end{equation}
and the rescaled quantity $a^2(\text{Var}(y') - 1)$ used as an input feature isolates the amplified noise components:
\begin{equation}
a^2(\text{Var}(y') - 1) \approx   \hat{t}_{\text{MLE}}^2 V_A + a^2 \hat{t^2} \hat \xi,
\end{equation}
since the discrepancy $(\hat{t} - \hat{t}_{\text{MLE}})^2$ vanishes in the large-sample limit due to the consistency of the MLE. Thus, the covariance term $\text{Cov}(x, y')$ is computed from the sample using the standard Pearson correlation estimator. The output of the network is then post-processed by dividing by $a^2$, recovering an accurate estimate of the original parameter $\hat{t}^2 \hat{\xi}$ from the amplified noise features.

\lqg{The use of an amplification factor $a > 1$ to rescale input features is a well-established preprocessing technique in neural networks. However, its value must be chosen with physical insight to meaningfully amplify the parameter of interest while preserving numerical stability. In our context, $a$ scales the excess-noise contribution in the input feature $a^2(\text{Var}(y')-1)$, making the excess noise more discernible against other signal components. The signal-to-noise ratio (SNR) of this amplified feature can be expressed as

\begin{equation}
\mathrm{SNR} = 10 \log_{10}{\frac{T V_\mathrm{A}}{\mu + a^2 T\xi}},
\end{equation}

For challenging regimes such as low transmittance ($T = 0.1$) and high excess noise ($\xi = 0.01$~SNU), the choice $a = 10$ with $V_A = 5 \ \mathrm{SNU}$ yields $\mathrm{SNR} \approx -3.42$~dB, a value within a operational range of communication systems. Hence, $a = 10$ represents a balanced selection that effectively elevates the excess-noise signal while maintaining estimation robustness across the considered parameter range.}

\subsection{Network Architecture}

\begin{figure*}
    \centering
    \includegraphics[width=0.9\linewidth]{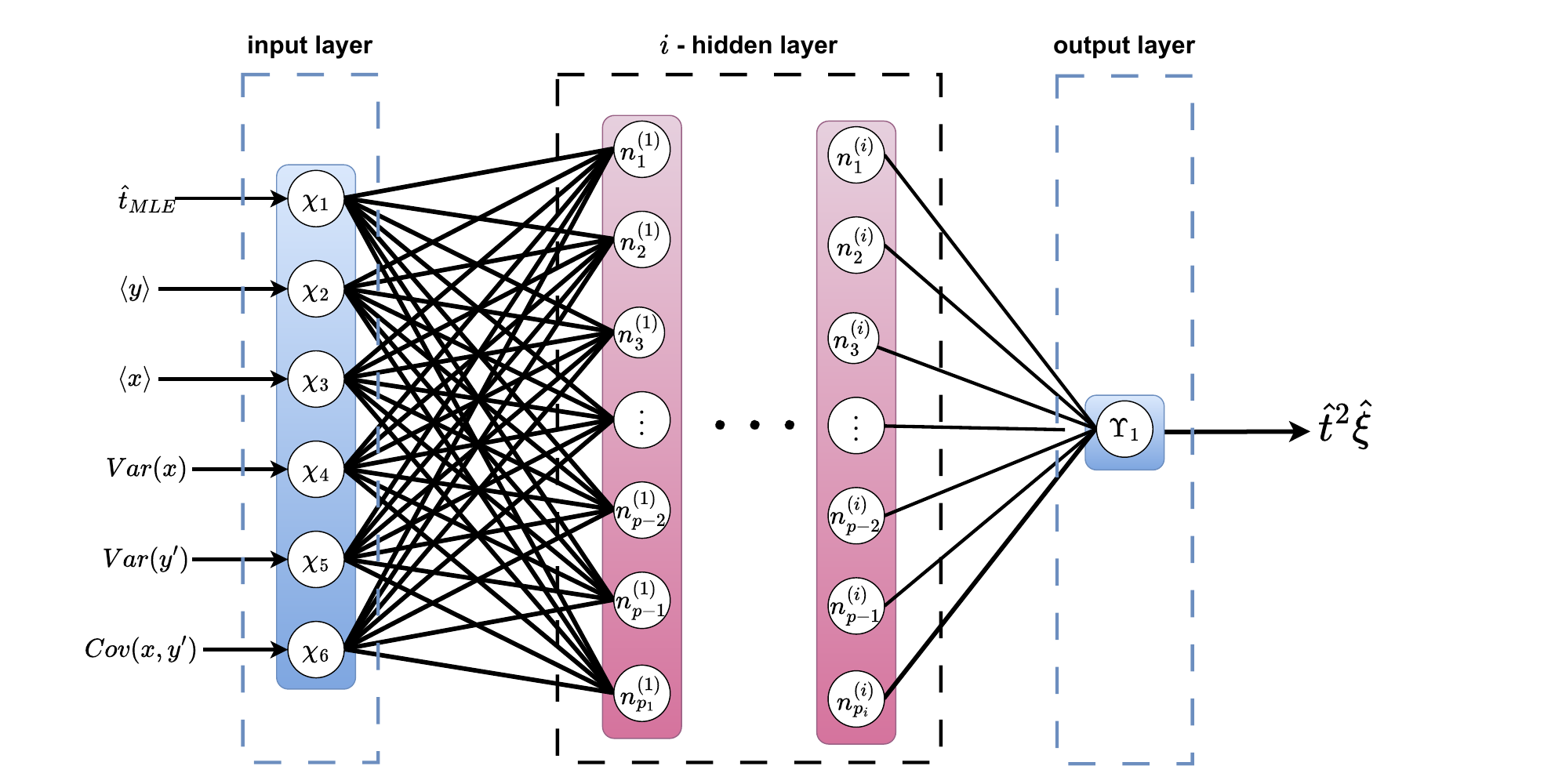}
    \caption{Neural network architecture for estimating the scaled excess noise \( t^2 \xi \).}
    \label{fig:network_arch}
\end{figure*}

The architecture of the network is illustrated in Fig. ~\ref{fig:network_arch}. The main goal here was to test the framework using a simple architecture, which comprises:

\begin{itemize}
    \item An input layer with six entry points, each corresponding to one of the features in \(\mathrm{X}_i\).
    \item A first hidden layer with 32 neurons using ReLU (Rectified Linear Unit) activation.
    \item A second hidden layer with 64 neurons, also using ReLU activation.
    \item A third hidden layer with 32 neurons without an explicit activation function prior to the final output transformation.
    \item An output layer consisting of a single neuron, whose output is passed through a shifted Softplus activation function, defined as:
    \begin{equation}
    \hat{\mathrm{Y}} = \log(1 + e^{z + b}),
    \end{equation}
    where \( z \) is the output of the final hidden layer and \( b \in \mathbb{R} \) is a learnable bias parameter initialized with a small positive value (\( b = 0.1 \)) to encourage strictly positive predictions.
\end{itemize}

This configuration was chosen to balance expressive power with simplicity, aiming for potential implementation on low-power embedded hardware. Accordingly, the neural network was designed with $p = 4450$ parameters, a feasible value for computing $\mathbf{F}(\hat \theta)$ (see Eq. \eqref{eq:jac}). The choice of ReLU activation potentially facilitates sparse activations and accelerates convergence, while the Softplus output ensures smooth nonlinearity and positivity, both properties desirable in the estimation of variance-like quantities.

\subsection{Training Strategy}

\label{sec:training}


The neural network is trained using supervised learning, where the target values are derived from synthetic data generated via a gaussian channel model (see Eq.~\eqref{eq:channel}). \lqgrev{At each step of the training loop, a new batch of data is produced by drawing the channel parameters uniformly at random: the transmittance \(t \in [0.01, 1]\) and the excess noise \(\xi \in [0.0, 0.1]\)~SNU. Specifically, 64 distinct transmittance values are drawn per loss evaluation, each paired with a random excess noise value, and this process repeats until convergence. This procedure ensures that the network is exposed to a continuously refreshed and representative sample of the channel parameter space, promoting robust generalization without overfitting to a fixed dataset.}

The loss function is defined as the mean squared error, ensuring the minimization of Eq.~\eqref{eq:min_loss} and consistency with the statistical framework introduced in Sec.~\ref{Worst-case}. The model parameters are optimized using the AdamW optimizer~\cite{loshchilov2019decoupledweightdecayregularization} implemented via the Optax library, leveraging the JAX and Flax frameworks for high-performance computation. Further details regarding the optimization procedure and hyperparameter settings are provided in Appendix~\ref{ap:OptimizationDetails}.

\section{Numerical investigations}

We investigate a finite-size security analysis employing neural networks within a specific architecture that leverages the signals $\{y_i\}_m$ and $\{x_i\}_m$ required for parameter estimation. A total of $10^5$ transmittance values $t$ are sampled, each associated with a corresponding noise variance $\sigma^2 = 1 + t^2 \xi$. For every sampled pair, we generate computationally $N$ signal using the discussed protocol. Channel parameters are then estimated using $m \equiv N/2$ signals in the PEP~\ref{PAP:2} and PEP~\ref{PAP:3}, which implement MLE and a neural network-based approach, respectively. This results in $10^5$ estimates for both $\sigma^2_{\text{max-MLE}}$ and $\sigma^2_{\text{max-NN}}$.

\lqg{The synthetic data used in our numerical investigations are generated according to the AWGN model (Eq.~\eqref{eq:channel}). This approach is standard for finite-size security analysis, as the gaussian attack represents the worst-case scenario for CV-QKD under the assumed channel model \cite{PhysRevLett.96.080502, Jain2022}. By using synthetic data, we maintain full control over channel parameters, enabling a precise assessment of the neural-network estimator's statistical performance under the most adversarial conditions. While experimental data could introduce additional non‑Gaussian imperfections, the synthetic‑data approach provides a fundamental validation of the estimator’s statistical properties and its compliance with finite‑size security proofs.}

\begin{figure}
    \centering
    \includegraphics[width=1.0\linewidth]{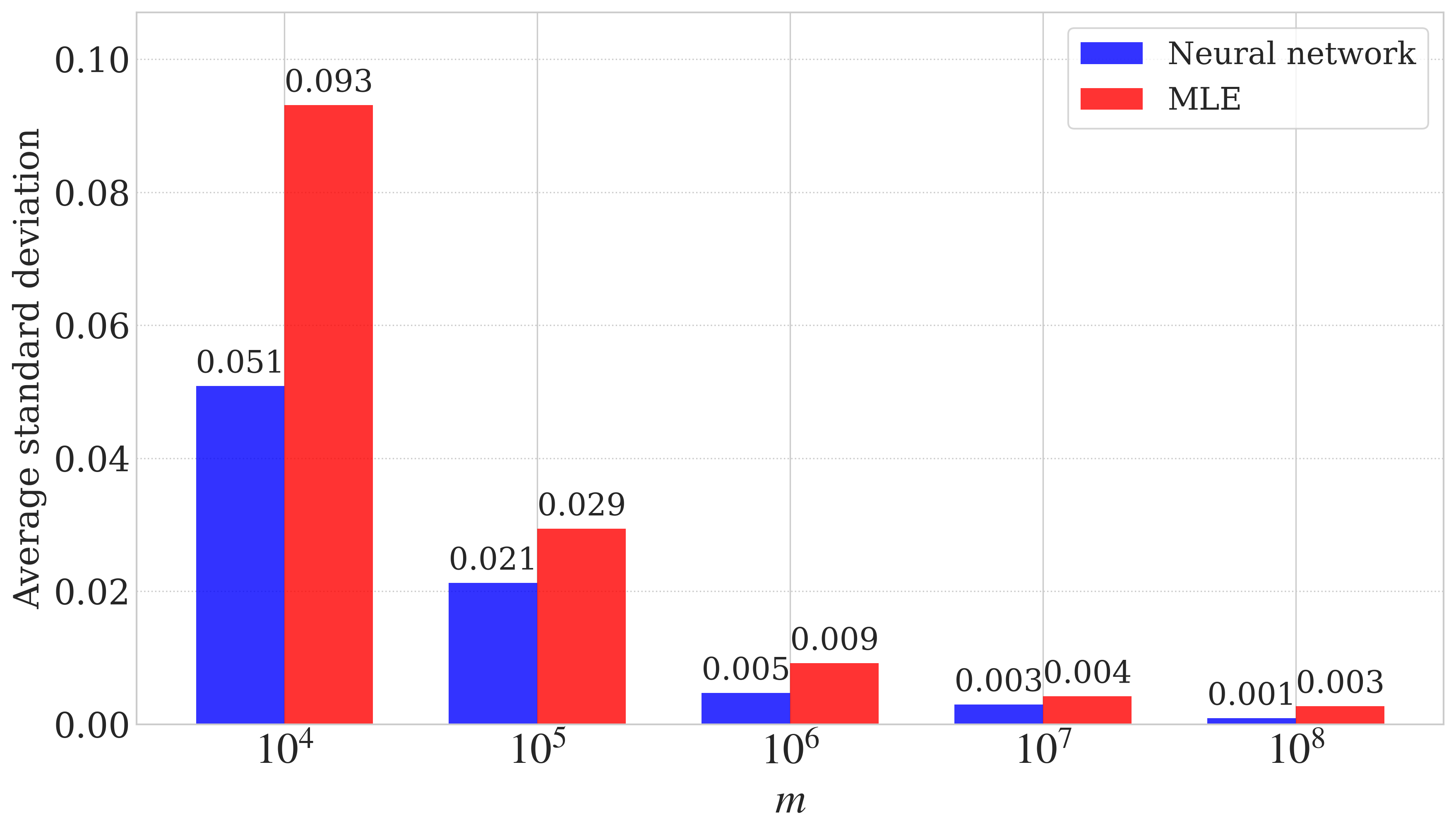}
    \caption{\justifying Standard deviation between the estimated channel parameters $\sigma^2_{\text{max-NN}}$ and $\sigma^2_{\text{max-MLE}}$ and the real values $\sigma^2$, using $m = 10^4$, $10^5$, $10^6$, $10^7$ and $10^8$ signals. The curves show that the average distance between the neural network estimation and the real values is smaller, which implies more precise estimations if compared to standard MLE method. }
    \label{fig:std_NN-MLE}
\end{figure}

As an initial benchmark, we compare the precision of the estimators by analyzing the standard deviation between the predicted and true values of $\sigma^2$. As shown in Fig.~\ref{fig:std_NN-MLE}, the neural network consistently yields lower deviations than the conventional MLE, reflecting the effectiveness of the error minimization performed during training. This increased accuracy stems from the fact that the network parameters are optimized to reduce a cost function (Eq. \eqref{eq:min_loss}), which directly penalizes large prediction errors. Furthermore, one can verify that $\sigma^2_{\text{max}} \rightarrow \sigma^2$ as the sample size $m$ increases, as expected from asymptotic consistency of both estimators (see Eq.~\eqref{eq:sigma_max} and Eq.~\eqref{eq:sigma_max_nn}).

\begin{figure}
    \centering
    \includegraphics[width=1\linewidth]{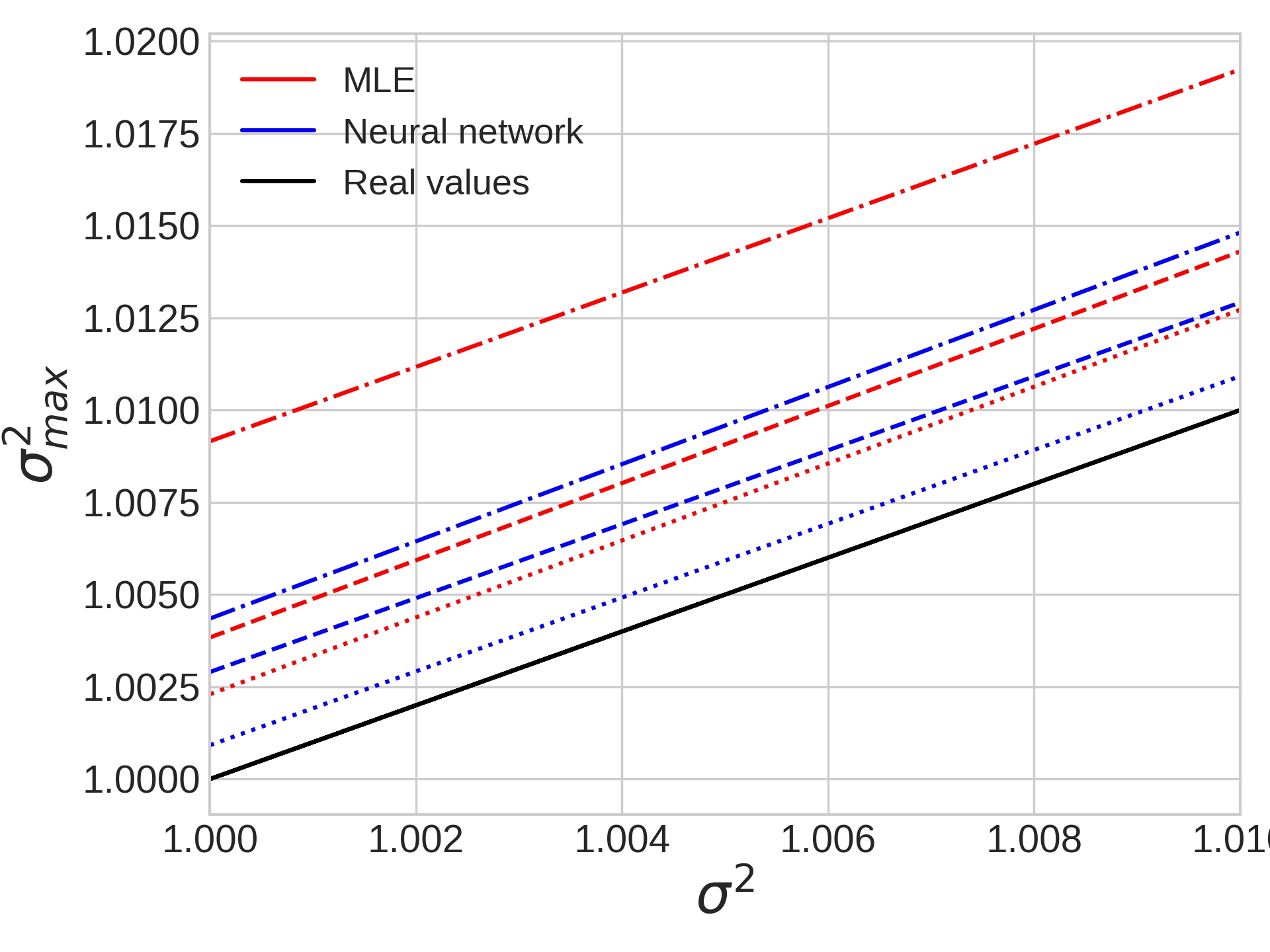}
    \caption{\justifying Comparison between the estimated and real values of $\sigma^2$. Dot-dashed line, dashed line and dotted line corresponds, respectively, to $m =10^6$, $10^7$ and $10^8$ signals. In all cases, the estimated values was never inferior to the real values.}
    \label{fig:m_std}
\end{figure}

However, the main challenge is not simply to show that neural networks can be more precise (a result already discussed in literature \cite{PhysRevA.97.022316, Chin2021, Luo_2022}), but demonstrate that they can also be $\epsilon_{PE}\text{-secure}$ for parameter estimation in CV-QKD. An estimate is $\epsilon_{PE}\text{-secure}$ if, and only if, all the points estimated are inside the confidence intervals with probability at least $1 - \epsilon_{PE}/2$, i.e., one can never estimate $\sigma^2_{\text{max}}$ below its real value considering this probability. Figure~\ref{fig:m_std} illustrates this behavior by depicting the average trend line obtained from the closest predicted points to the ideal reference, computed across all samples. \lqg{The results empirically validates this operational equivalence, demonstrating that, for all tested sample sizes, the neural estimate $\sigma^2_{\text{max-NN}}$ never falls below the true value $\sigma^2$, thereby satisfying the security condition with the specified probability $\epsilon_{PE}$.}


To analyze the impact of improved parameter estimation on the secret-key rate, we set $p_{ec} = 0.9$ for the probability of successful error correction and adopt $\epsilon_{PE} = \epsilon_{\text{cor}} = \bar \epsilon = \epsilon_{PA} = 10^{-10}$ for the security parameters, yielding an overall composable security level of $\epsilon \approx 3.9 \cdot 10^{-10}$ against collective Gaussian attacks, as described in Eq.~\eqref{eq:skr_finite}. The parameters summarized in Tab. ~\ref{tab:parameters} are selected to reflect realistic conditions, based on experimental implementations reported in the literature \cite{10.1063/5.0179566, Lu2019, Milicevic2018, Chin2021, Jain2022}, aiming to ensure practical feasibility.

\begin{table}
\centering
\caption{\justifying Protocol parameters used in this work. The parameter $d$ is chosen based on ref. \cite{Jain2022}. Detector efficiency values reflect specifications of commercially available detectors \cite{10.1063/5.0179566, Lu2019}. The reconciliation efficiency and probability of success of error correction is set according to recent experimental implementations \cite{Milicevic2018, Chin2021, Jain2022}.}

\begin{tabular}{l@{\hskip 1cm}l@{\hskip 1cm}l} 
\hline
\textbf{Protocol parameter} & \textbf{Symbol} & \textbf{Value} \\
\hline
discretization & $d$ & 32 \\
Quantum duty & $\mu$ & 1 (hom.) \\
Detector efficiency & $\eta_{\mathrm{eff}}$ & 0.8  \\
Excess noise & $\xi$ & $0.01$ SNU \\
Variance & $V_A$ & $5$ SNU \\
Left signals fraction & $n/N$ & $0.5$ \\
Reconciliation efficiency & $\beta$ & $0.95$ \\
Probability of success & $p_{ec}$ & $0.9$ \\
of error correction & & \\
\hline
\label{tab:parameters}
\end{tabular}
\end{table}

\begin{figure*}
    \centering
    \includegraphics[width=0.7\linewidth]{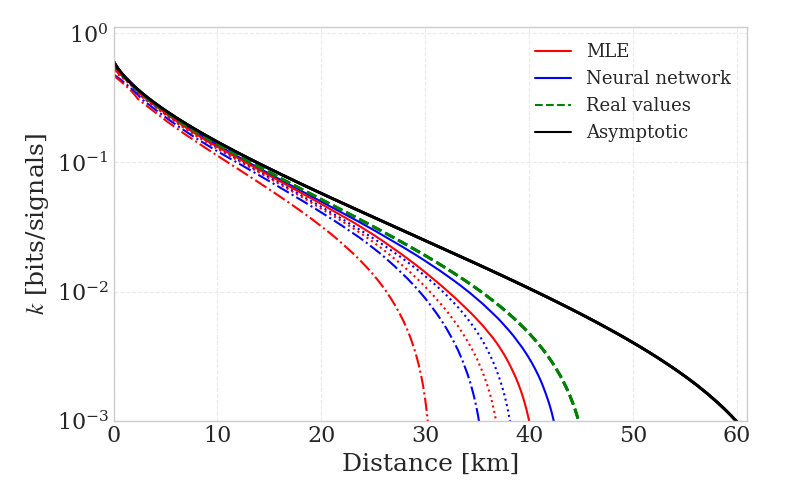}
    \caption{\justifying Secret-key rate using the discussed protocol with parameters described in Tab. \ref{tab:parameters}. Dot-dashed line, dashed line and dotted line corresponds, respectively, to $N = 2 \cdot 10^6$, $2 \cdot 10^7$ and $2 \cdot 10^8$ signals. The use of neural networks allowed a gain of $6.1$ km, $1.3$ km and $3.0$ km, respectively. In all cases, the estimated values was never superior to the real values.}
    \label{fig:skr_km}
\end{figure*}

The precision of parameter estimation plays a critical role in the secret-key rate — since we must overestimate the excess noise with high confidence, more accurate estimators yield smaller values of $\xi$, ensuring that $\Gamma_{\epsilon_{PE}} \in \mathcal{C}_{\epsilon_{PE}}$. The results in Fig.~\ref{fig:skr_km} illustrate this behavior: the estimated secret-key rate consistently remains below the ideal rate, as expected. This outcome confirms the operational security of the parameter estimation procedure described in PEP~\ref{PAP:3}, with the advantage of achieving higher rates.

Although the computational cost during training is considerable, inference is highly efficient and readily implementable in practical scenarios. This balance is particularly critical in the finite-size regime with limited signals, where even small improvements in parameter estimation can significantly extend the achievable communication distance between Alice and Bob, enabling secure key distribution in conditions where traditional estimators fall short.

\section{Conclusion}

\lqg{Neural networks are increasingly being adopted as estimation tools in quantum information, with their application in QKD highlighting a critical balance between computational cost and estimation accuracy. While these models typically require greater resources than conventional methods, they can achieve substantially improved precision. This enhancement is particularly significant in the finite size regime, where even modest gains in parameter estimation can enable a positive secret key rate in otherwise insecure operational regimes. The neural network estimator presented here demonstrates how machine learning can be securely integrated to enhance practical QKD performance. By learning subtle statistical dependencies within correlated quadrature data, the network delivers tighter bounds on excess noise than standard linear estimators, directly leading to higher secret-key rates and extended transmission distances. This improvement is especially valuable when the number of transmitted signals is limited, as it can reduce the parameter estimation overhead and extend the secure range of existing systems.}

\lqg{The developed framework is broadly applicable to Gaussian modulated CV QKD under collective attacks and paves the way for future adaptations, including discrete modulation, time varying channels, or hybrid schemes incorporating real time feedback for adaptive estimation. Considering our approach, the neural network estimator provides tighter parameter bounds for collective Gaussian attacks, which form the basis for security against coherent attacks via reduction techniques. Since these reductions rely only on the validity of the worst-case bounds, our method can be seamlessly integrated into composable security proofs against general attacks. Thus, the gains in key rate and distance demonstrated in this work are preserved while maintaining the same security level against coherent attacks.}



In conclusion, this article provides a finite-size analysis for secure CV-QKD using networks for excess noise estimation. While the neural network employed in our simulations demonstrates improved estimation accuracy compared to the conventional MLE, we emphasize that the primary contribution of this work lies not in the superiority of a specific architecture, but in demonstrating that neural network-based estimators can be incorporated into parameter estimation routines for CV-QKD without compromising composable security. Although more robust or efficient architectures may be developed, our findings indicate that such data-driven approaches are compatible with finite-size security proofs. This insight enables the use of flexible and potentially adaptive estimation strategies in practical QKD systems, paving the way for further integration of machine learning techniques into secure quantum communication protocols.

\begin{acknowledgments}
This work has been fully funded by the project Computational Architecture for Flexible QKD System Post-processing Platform and \lqgrev{QRNG: Post-processing and Experimental} supported by QuIIN - Quantum Industrial Innovation, EMBRAPII CIMATEC Competence Center in Quantum Technologies, with financial resources from the PPI IoT/Manufatura 4.0 of the MCTI grant number 053/2023, signed with EMBRAPII. \lqgrev{MD thanks financing from the European Union (HORIZON-MSCA-2023 Postdoctoral Fellowship, 101153602 - COCoVaQ)}.

\end{acknowledgments}

\appendix

\section{Computational cost of the neural network}

\label{ap:compcost}

The computational complexity of the neural architecture is primarily determined by the number of trainable parameters and the per-sample inference cost. The model is implemented in Flax and trained using the JAX framework, leveraging XLA compilation and hardware acceleration for efficient execution. Let $n$ be the number of samples per iteration and $d$ the input dimension, with $d = \lqg{6}$ corresponding to the MLE estimate, mean values, variances, and covariances extracted from the AWGN model. The network processes input tensors of shape $(n, d)$.

Assuming a fully connected feedforward neural network with \( L \) layers and \( h \) hidden units per layer, the time complexity of a forward or backward pass is approximated by:

\begin{equation}
\mathcal{O}(d h + (L - 1) h^2) \approx \mathcal{O}(L d h + L h^2)\,.
\end{equation}

The term \( \mathcal{O}(d h) \) corresponds to the affine transformation from the input to the first hidden layer, while \( \mathcal{O}((L - 1) h^2) \) results from the matrix multiplications between subsequent hidden layers \cite{goodfellow2016deep}.

Training and inference benefit from JAX primitives such as \texttt{jit} and \texttt{vmap}, enabling automatic parallelization and just-in-time compilation. Pseudo-random number generation for AWGN simulations is handled deterministically via key splitting to ensure reproducibility.

\lqg{\section{Practical implementation for confidence interval computation}
\label{ap:practical}

\textit{Jacobian computation for varying sample sizes.}
In the numerical investigations, computing the confidence interval in Eq.~\eqref{eq:conf_int_nn} requires the Jacobian matrix \(\mathbf{F}(\hat{\theta})\). Rather than recalculating the full \(n \times p\) Jacobian \lqgrev{ and the $s$ variance} for each sample size \(m\), which would be computationally expensive, we adopt an efficient practical approach. After training and fixing the network parameters \(\hat{\theta}\), we generate one large synthetic dataset with \(m_{\max} = 10^8\) samples \lqgrev{considering the approach described in Sec. \ref{sec:training}}. For any smaller sample size \(m\), we then use only the first \(m\) rows of this precomputed Jacobian. This method is statistically valid because the data are i.i.d. and the network parameters remain unchanged after training. It reduces the computational overhead during inference to a single Jacobian evaluation, making the procedure feasible across the range of sample sizes considered.


\textit{Regularization of the matrix inversion.}
In practice, the matrix \(\mathbf{F}^T\mathbf{F}\) is often ill-conditioned, especially when the number of samples \(m\) is not substantially larger than the number of parameters \(p\). \lqgrev{The observed ill‑conditioning arises because a significant proportion of its eigenvalues lie extremely close to zero. In practice, finite‑precision arithmetic can produce a small number of eigenvalues with a negligible negative magnitude (\(\sim 10^{-10}\)), which we treat as effectively zero. The presence of these near‑zero eigenvalues renders the matrix nearly singular and causes its inversion to be numerically unstable.}

To ensure numerical stability when computing the inverse \((\mathbf{F}^T\mathbf{F})^{-1}\), we employ Tikhonov regularization. Specifically, we replace \((\mathbf{F}^T\mathbf{F})^{-1}\) with \((\mathbf{F}^T\mathbf{F} + \lambda \mathbf{I})^{-1}\), where \(\lambda > 0\) is a small regularization parameter and \(\mathbf{I}\) is the \(p \times p\) identity matrix. For all simulations reported in this work, we set \(\lambda = 10^{-4}\)\lqgrev{, resulting in a condition number $\kappa \approx 4.5 \cdot 10^3$.} This value was chosen via cross-validation on a separate validation set, aiming to minimize the mean squared error of the variance predictions while maintaining well-conditioned matrix inversions. \lqgrev{Therefore, the regularization transforms the original ill-conditioned matrix $\mathbf{F}^T\mathbf{F}$ (with $\sim 15\%$ of its eigenvalues below $\lambda$) into the regularized matrix $\mathbf{F}^T\mathbf{F} + \lambda \mathbf{I}$, which has all eigenvalues $\geq \lambda$, eliminating any numerical instability and  practical rank deficiency arising from near-zero eigenvalues.}  


}

\djgs{
\section{Inference latency and real‑time feasibility}

To generate Mbit/s of security keys, a practical real‑time CV‑QKD system should be able to process tens or hundreds of MBaud ~\cite{10.1063/5.0179566}. In this context, parameter estimation must be completed within stringent real‑time limits, typically processing a portion of that system throughput. The neural network estimator introduced in this work would be able to meets this requirement by operating exclusively in inference mode once offline training is completed.

A single forward pass determines its computational demand through a fully‑connected network with $p = 4450$ parameters, which translates to $8679$ elementary arithmetic operations per frame. This computational cost is primarily driven by matrix‑vector multiplications in the hidden layers. It is constant and does not scale with the number of symbols processed per frame, being negligible compared to the required data preprocessing to calculate its inputs~\eqref{eq:nnpe-inputs}, which include the statistical moments and the transmittance MLE~\eqref{eq:MLE}.

Consequently, the computational burden is dominated by the linear complexity $\mathcal{O}(m)$ required to prepare the statistical inputs. Both the conventional analytic method and the neural-network approach must first compute $\hat{t}_{MLE}$, a step whose cost is common to both. The divergence occurs in the estimation of the variance and excess noise. The classical method computes the sum of squared residuals, iterating once over the data with the already obtained $\hat{t}_{MLE}$. The neural-network method, in turn, requires the calculation of five sufficient statistics (means, variances, and a covariance) from the raw data vectors. As detailed in Table~\ref{tab:comp_cost}, the cost of generating these statistical inputs for a frame of $m=10^8$ symbols is linear and comparable between the two methods. The key distinction emerges after this stage: while the analytic method proceeds with a final arithmetic step, the neural-network approach performs a single forward pass through a fixed architecture. This inference requires only $\approx 8.7 \times 10^3$ operations, which is a negligible, constant overhead that is effectively imperceptible against the $\mathcal{O}(10^8)$ operations already expended on the shared transmittance estimation and statistical aggregation.

\begin{table}[h] \lqg{
\centering
\begin{threeparttable}
\caption{\justifying \lqg{Computational cost per frame for parameter estimation with $m = 10^8$ correlated measurement pairs. The table compares the dominant, shared cost of transmittance estimation with the subsequent steps for noise variance estimation, highlighting the fixed and negligible overhead introduced by the neural-network inference. All costs are estimated in terms of arithmetic operations.} }
\label{tab:comp_cost}
\begin{tabular}{l r}
\hline
\textbf{Processing Stage \& Scope} & \textbf{Est. Cost (Ops)} \\
\hline
\multicolumn{2}{l}{\textit{\textbf{Transmittance Estimation (Shared Input)}}} \\
Calculation of $\hat{t}_{MLE}$ & $\mathcal{O}(m) \approx 2 \times 10^8$ \\
\quad Divisions* & $+ 1 \times 10^8$ \\
\hline
\multicolumn{2}{l}{\textit{\textbf{Variance Estimation}}} \\
\textbf{Analytic (MLE) Approach:} & \\
\quad Residual computation $\sum (y_i - \hat{t}x_i)^2$ & $\mathcal{O}(m) \approx 4.0 \times 10^8$ \\
\cline{1-1}
\textbf{Neural-Network-Based Approach:} & \\
\quad Preparation of 5 sufficient statistics & $\mathcal{O}(m) \approx 8.0 \times 10^8$ \\
\quad Neural-network inference (single forward) & $\mathbf{\mathcal{O}(1) \approx 8.7 \times 10^3}$ \\
\hline
\end{tabular}
\begin{tablenotes}
      \small
      \item[*] Division operations are computationally expensive and can take many cycles depending on the hardware support.
\end{tablenotes}
\end{threeparttable}}
\end{table}

In terms of execution time, a single forward pass executes well below one millisecond on a standard CPU and can be accelerated to microsecond latency when deployed on embedded processors, FPGAs, or dedicated digital signal processing platforms. In a test performed on an NVIDIA RTX3090 GPU, with $10000$ runs, the average inference time was $329.65 \mu s$, with a standard deviation of $40.20 \mu s$.

Notably, no online training or backpropagation is performed during operation. Furthermore, the Jacobian matrix required for constructing the worst‑case confidence interval is pre‑computed offline using a large synthetic dataset and reused at runtime. These characteristics confirm that the neural‑network estimator is not only statistically sound but also practically viable for integration into real‑world CV‑QKD systems, where low‑latency and deterministic processing are essential.

}

\djgs{

\section{Optimization details}
\label{ap:OptimizationDetails}

The network parameters were optimized using the AdamW optimizer~\cite{loshchilov2019decoupledweightdecayregularization}, implemented via the Optax library within the JAX and Flax frameworks. An exponential learning-rate decay schedule was employed, with an initial learning rate of $1\times10^{-4}$, transition steps of 500, and a decay rate of 0.9. The weight decay coefficient was set to $1\times10^{-5}$.

Training was performed using a batch size of 64 samples for 15 epochs, with 512 gradient-update steps per epoch. For numerical stability, gradients were clipped by global norm with a threshold of 0.5. Model performance was periodically evaluated every 256 training steps. No adaptive hyperparameter tuning, learning-rate warm-up, or early stopping strategy was applied. All experiments were conducted using a fixed JAX pseudo-random number generator seed to ensure reproducibility.
}

\bibliography{sample}

\end{document}